%% file: main.tex
\documentclass[conference]{IEEEtran}
\usepackage{cite}
\usepackage{amsmath,amssymb,amsfonts}
\usepackage{algorithmic}
\usepackage{graphicx}
\usepackage{textcomp}
\usepackage{xcolor}
\usepackage[a4paper, total={184mm,239mm}]{geometry}
\usepackage{booktabs}
\usepackage{multirow}
\usepackage{amssymb}
\usepackage{array}
\usepackage{subcaption}
\usepackage{hyperref}

\usepackage[dvipsnames]{xcolor}
\usepackage{xspace}
\usepackage{longtable}
\usepackage{tabularx}
\usepackage{enumitem}

\usepackage{array,booktabs}
\newcolumntype{R}{@{}r@{}}
\newcommand{\sep}[1]{\multicolumn{1}{@{}c@{}}{\;#1\;}}
\newcolumntype{N}{@{}c@{}}

\begin{document}

\newcommand{\JZ}[1]{\textcolor{blue}{#1}}
\newcommand{\comment}[1]{\textcolor{orange}{#1}}
\newcommand{\todo}[1]{\textcolor{red}{#1}}

\newcommand{\yes}{\textcolor{Green}{\checkmark}}
\newcommand{\no}{\textcolor{red}{$\times$}}

\title{
DreamRAM: A Fine-Grained Configurable Design Space Modeling Tool for Custom 3D Die-Stacked DRAM
}

\input{01authors.tex}
\maketitle

\input{02abstract.tex}

\input{03introduction.tex}

\input{05methodology.tex}

\input{04background.tex}

\input{06results.tex}

\input{07relatedworks.tex}

\input{08conclusion}

\bibliographystyle{IEEEtran}
\bibliography{99references}
\clearpage
\input{88appendix}

\end{document}

%% file: 01authors.tex
\author{\IEEEauthorblockN{Victor Cai$^\dag$, Jennifer Zhou$^\ddagger$, Haebin Do$^\ddagger$, David Brooks$^\ddagger$, and Gu-Yeon Wei$^\ddagger$}
\IEEEauthorblockA{Harvard University}
\IEEEauthorblockA{\footnotesize $^\dag$victorcai@college.harvard.edu $^\ddagger$\{jennifer\_zhou, haebin\_do, dbrooks, guyeon\}@g.harvard.edu}
}

%% file: 02abstract.tex
\begin{abstract}

3D die-stacked DRAM has emerged as a key technology for delivering high bandwidth and high density for applications such as high-performance computing, graphics, and machine learning. However, different applications place diverse and sometimes diverging demands on power, performance, and area that cannot be universally satisfied with fixed commodity DRAM designs. Die stacking creates the opportunity for a large DRAM design space through 3D integration and expanded total die area. To open and navigate this expansive design space of customized memory architectures that cater to application-specific needs, we introduce DreamRAM, a configurable bandwidth, capacity, energy, latency, and area modeling tool for custom 3D die-stacked DRAM designs. DreamRAM exposes fine-grained design customization parameters at the MAT, subarray, bank, and inter-bank levels, including extensions of partial page and subarray parallelism proposals found in the literature, to open a large previously-unexplored design space. DreamRAM analytically models wire pitch, width, length, capacitance, and scaling parameters to capture the performance tradeoffs of physical layout and routing design choices. Routing awareness enables DreamRAM to model a custom MAT-level routing scheme, Dataline-Over-MAT (DLOMAT), to facilitate better bandwidth tradeoffs. DreamRAM is calibrated and validated against published industry HBM3 and HBM2E designs. Within DreamRAM's rich design space, we identify designs that achieve each of 66\% higher bandwidth, 100\% higher capacity, and 45\% lower power and energy per bit compared to the baseline design, each on an iso-bandwidth, iso-capacity, and iso-power basis. 

\end{abstract}

%% file: 03introduction.tex
\section{Introduction}

Historically, DRAM design has taken capacity and cost as top priorities.  Over the decades, several commodity DRAM families emerged to serve different markets: DDR for general-purpose computing, LPDDR for mobile and embedded platforms, GDDR for graphics, and High Bandwidth Memory (HBM) for data-intensive workloads. With the exception of HBM, these families were designed around large-volume adoption, making density the dominant axis of optimization. 
\begin{figure}[!htbp]
\centering
\includegraphics[width=1\linewidth]{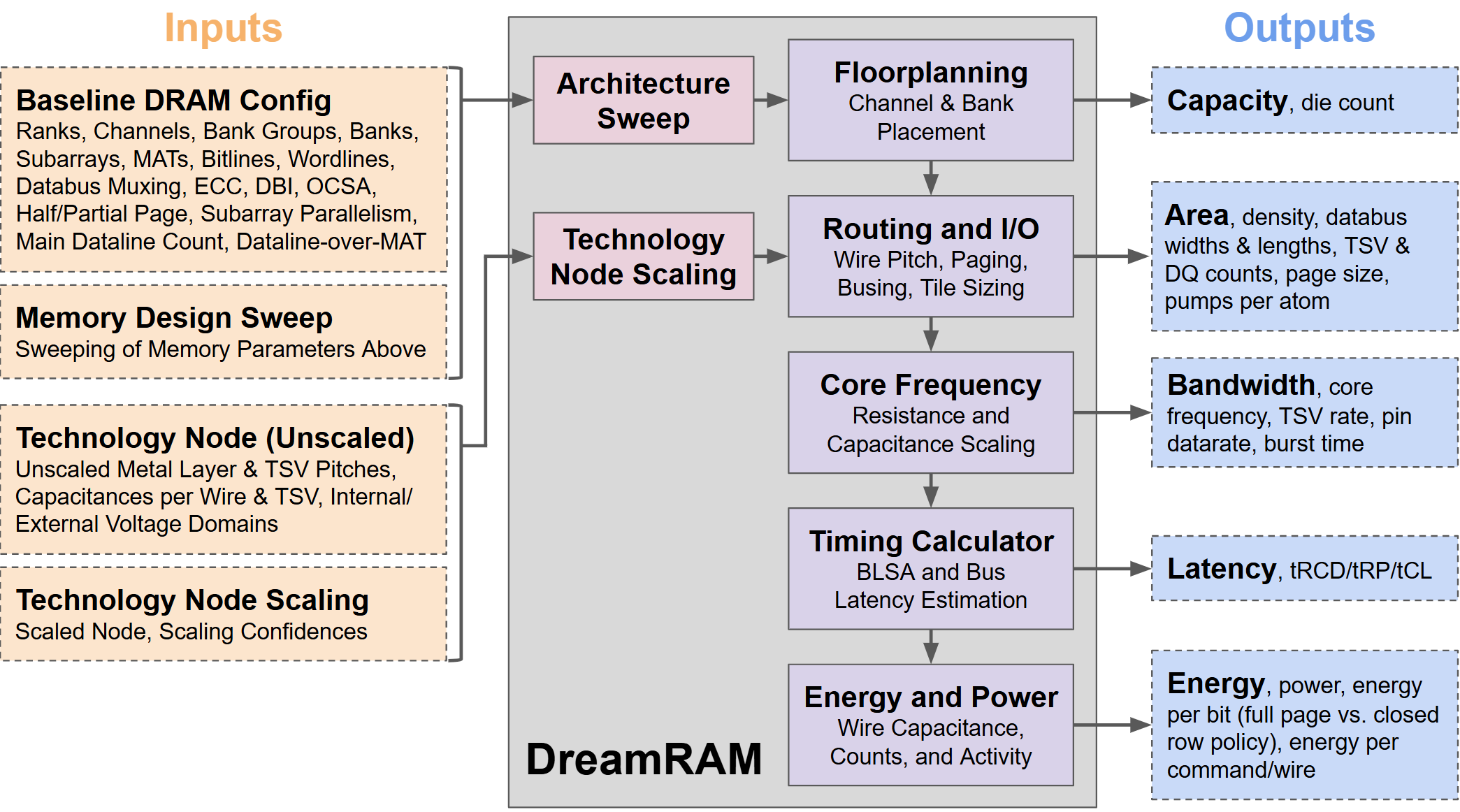}
\caption{The DreamRAM framework.}
\label{fig:framework}
\end{figure}
With HBM, DRAM design experienced a paradigm shift. Metrics such as bandwidth and energy efficiency have been elevated to first-order design goals. This shift was driven by changes in application requirements: modern workloads in high-performance computing, graphics, and machine learning place tremendous pressure on memory systems, and consumers are willing to pay for hardware solutions that deliver performance beyond what density-driven designs can provide, rather than treating DRAM solely as a commodity product. HBM successfully demonstrated the potential of specialized memory systems tailored towards application-specific demands. However, today’s DRAM design space and simulators remain constrained (see related works in Section \ref{sec:relatedworks}.) No framework allows systematic analysis of tradeoffs across the broad design space of custom DRAM beyond the limits of standard commodity designs.

To address this gap, we propose \textbf{\textit{DreamRAM}}, a configurable modeling tool for custom 3D die-stacked DRAM architectures. DreamRAM analytically models bandwidth, capacity, energy, latency, and area while exposing fine-grained design parameters at the MAT, subarray, bank, and inter-bank levels. By providing a unified and extensible exploration framework, DreamRAM enables researchers and designers to uncover new opportunities for workload-tailored memory design.

\begin{figure*}[!t]
\centering
\includegraphics[width=0.9\linewidth]{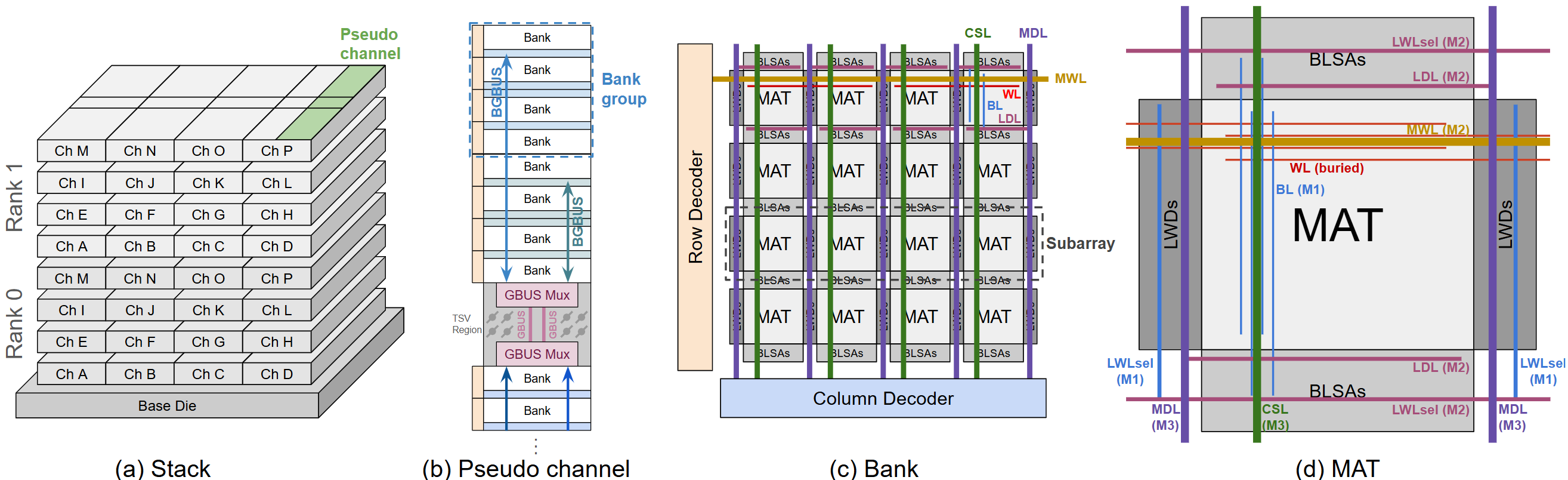}
\caption{HBM3 organization at the (a) stack, (b) inter-bank, (c) bank/subarray, and (d) MAT levels.}
\label{fig:background}
\end{figure*}

This paper makes the following contributions: 
\begin{itemize}
    \item We introduce a parameterized 3D die-stacked DRAM modeling tool, \textbf{\textit{DreamRAM}}, that exposes a range of design knobs at the MAT, subarray, bank, and inter-bank levels. We incorporate DRAM modifications for partial pages and subarray parallelism \cite{chatterjee_architecting_2017}\cite{ha_improving_2016}\cite{kim_case_2012}.
    \item We model the routing pitch, length, capacitance, and scaling of wires down to the MAT level. As an example, we use this routing awareness to model a new MAT-level routing scheme, Dataline-over-MAT (DLOMAT). 
    \item We demonstrate a vast 3D die-stacked DRAM design space in terms of bandwidth, capacity, energy, latency, and area. We illustrate how enabling more fine-grained design parameters significantly increases the design space range and highlight design choices that optimize DRAMs for each metric.
    \item We showcase how the DreamRAM design space enables designers to visualize the tradeoffs and constraints of metrics in different application scenarios (server CPU, server GPU, high-performance edge, and embedded IoT). As a further case study on server GPUs, we optimize bandwidth, capacity, and energy efficiency while enforcing constraints on the other metrics. 
    \item The DreamRAM simulator is open-source and can be found here: \url{https://github.com/harvard-acc/dreamram}
\end{itemize}

Section \ref{sec:methodology} describes the DreamRAM simulator framework. Section \ref{sec:background} outlines DreamRAM's DRAM modeling and parameterization. Section \ref{sec:results} presents validation and results. Finally, section \ref{sec:relatedworks} discusses related works. 

%% file: 05methodology.tex
\section{DreamRAM Framework}\label{sec:methodology}
Fig. \ref{fig:framework} summarizes the DreamRAM framework and details the input parameters, submodules, and outputs. DreamRAM takes four input files: a baseline memory configuration, a memory parameter sweep description, an unscaled technology node, and technology node scaling parameters. DreamRAM defaults to a baseline HBM3 configuration \cite{ryu_16_2023}, a 2ynm unscaled node \cite{ha_understanding_2018}, and a 1znm scaled node \cite{ryu_16_2023}. The baseline configuration parameterizes a single DRAM design, while the memory sweep defines sweep ranges for each memory parameter. Since node information is often unavailable, the node scaling file provides confidence inputs for how well unscaled parameters for capacitances, logic, sense amplifiers, and wordline drivers scale with technology node. A scaled technology node is calculated from the unscaled node and these scaling confidence parameters. 

\subsection{Outputs and Metrics}
DreamRAM's primary output metrics are capacity, area, bandwidth, latency, and energy per bit. \textbf{Capacity} depends only on the DRAM configuration. \textbf{Area} is calculated hierarchically from the bitlines/wordlines up to the full die. For circuit modifications, DreamRAM separately estimates the transistor overheads and datapath routing overheads, and takes the larger overhead. \textbf{Bandwidth} combines the input architecture with a bank cycle time estimate discussed along with \textbf{Latency} in Section \ref{sec:corefreq}. \textbf{Energy per bit} is found using Eq. \ref{eq:energy}, where $\alpha$ is an activity factor. DreamRAM also outputs power calculated as the product of bandwidth and energy per bit. 
\begin{align}\label{eq:energy}E=\sum\frac12 \alpha\; n\;(C_{per\;l}\;l)\;\Delta V_{internal}V_{external}\end{align}
DreamRAM's submodules for floorplanning, routing, core frequency, timing, and energy provide various related outputs as seen in Fig. \ref{fig:framework}. While cost remains an important factor in DRAM design, its dependence on yield, process maturity, and other fabrication-specific variables renders rigorous modeling beyond the scope of this work. We approximate cost to first order with area where appropriate.

%% file: 04background.tex
\newcolumntype{P}[1]{>{\centering\arraybackslash}p{#1}}

\section{DRAM Organization and Parameterization}\label{sec:background}
DRAM is organized hierarchically into the inter-bank, bank, subarray, and MAT levels. Inter-bank level parameters change the rank, channel, pseudo channel, and bankgroup structures but leave banks unchanged. Bank level refers to changes to row and column decoding that leave subarray or MAT patterns untouched. The subarray level allows changes to subarray wires but does not affect MAT operation. Finally, the MAT level can modify the MAT organization and associated routing. We break down DreamRAM's parameterization of DRAM at each level.

\subsection{Inter-Bank Organization}
A DRAM channel connects to independent command address (CA) and data (DQ) pins. A channel transfers data in units of 32-byte atoms. Ranks provide extra capacity by increasing the number of chips that share the same CA bus and DQs. In DDR, each chip drives a subset of the DQs, so each access is spread across all chips in the rank. Since DDR4, banks on a chip have been organized into bank groups that share global I/O routing. 

\textbf{3D Die-Stacked DRAM.}
HBM is enabled by through-silicon vias (TSVs) that run commands, data, and power between dies. An HBM stack has 1024 DQs divided among multiple channels. Fig. \ref{fig:background} (a) shows how the same channels of different ranks are vertically aligned across different dies to allow ranks to share TSVs for channel parallelism\cite{ke_chen_cacti-3dd_2012}, making HBM more than a blind stacking of 2D dies. Each access is routed to one bank on one die, greatly improving energy efficiency. HBM2 introduces pseudo channels, which share a common CA bus but maintain separate command decoders, banks, and DQ interfaces. As shown in Fig. \ref{fig:background} (b), bank data is muxed onto a bank group bus (BGBUS), then a global bus (GBUS) before descending through TSVs to the base die.

\textbf{DreamRAM Inter-Bank Parameterization.} At the inter-bank level, DreamRAM adopts a vendor-published HBM3 floorplan\cite{ryu_16_2023}\cite{park_192-gb_2022} as its modeling baseline. DreamRAM supports input parameters for the number of ranks, channels, channels/die, bank groups, and banks. To support earlier HBM designs \cite{chun_16-gb_2021}, bank group parameterization is split into horizontal and vertical tiling, with commands routed either between pseudo channels or within a pseudo channel between its bank groups. We model the muxing and speed of BGBUSes, GBUSes, TSVs, and DQs for the alternative data line (ADL) described in \cite{ryu_16_2023}. 

\subsection{Bank, Subarray, and MAT Organization}
A bank contains MATs grouped horizontally into subarrays, as in Fig. \ref{fig:background} (c) and (d). Horizontal wordlines (WLs) and vertical bitlines (BLs) intersect at the cells. Adjacent MATs share local WL drivers (LWDs) and BL sense amplifiers (BLSAs). During activation (ACT), the bank asserts a main wordline (MWL) and a local WL select (LWLsel)\cite{chatterjee_architecting_2017}, which raises WLs in the MATs to open a page. In the open bitline scheme, BLSAs amplify the BLs in the activated MAT and BLs in adjacent MATs to opposite logic levels. A read/write command fires column select lines (CSLs) in the MATs, propagating data onto the local and main datalines (LDLs \& MDLs). Precharge (PRE) resets all wordlines and bit/datalines for a new ACT. 

\input{tab-validation}
\input{tab-volume5d}

\textbf{DreamRAM Bank and Subarray Parameterization.}
At the bank level, DreamRAM exposes parameters for the number of subarrays, MATs per subarray, and repair subarrays. DreamRAM also parameterizes and incorporates several useful DRAM bank and subarray-level modifications from the literature. We model offset-cancellation sense amplifiers (OCSAs) \cite{kim_sensing_2019, marazzi_hifi-dram_2024} that can be instantiated in place of conventional BLSAs. We model Subarray-level Parallelism (SALP)\cite{kim_case_2012}, which modifies the bank row decoder to latch each subarray row address separately, allowing multiple subarrays to be active (but not concurrently accessed). SALP does not mitigate adjacent subarray conflicts in the open bitline scheme, where activating a subarray occupies bitlines in adjacent subarrays due to shared BLSAs. DreamRAM implements two solutions: (1) DreamRAM's SALP-groups parameter prevents conflict by inserting buffer subarrays between subarray groups\cite{ha_understanding_2018}, where each group can open one page, without as much area overhead as splitting out banks. (2) DreamRAM's SALP-all setting does not add buffers, leaving subarray conflict management to the memory controller, which can keep up to half the subarrays active. We also model strategies that reduce page activation size for energy efficiency by modifying row-address decoding, including half page \cite{ha_improving_2016} and subchannels \cite{chatterjee_architecting_2017}. We term these proposals ``partial page strategies.'' While \cite{ha_improving_2016} doubles the LDLs to maintain access to all the MDLs, \cite{chatterjee_architecting_2017} leaves each subchannel's subset of MDLs semi-independent, obtaining a full data atom from each partial page over multiple cycles. We generalize these cycles as ``pumps.'' We add a similar option for \cite{ha_improving_2016} to leave the MDL subsets similarly independent.

\begin{figure}[b]
    \centering
    \includegraphics[width=0.9\linewidth]{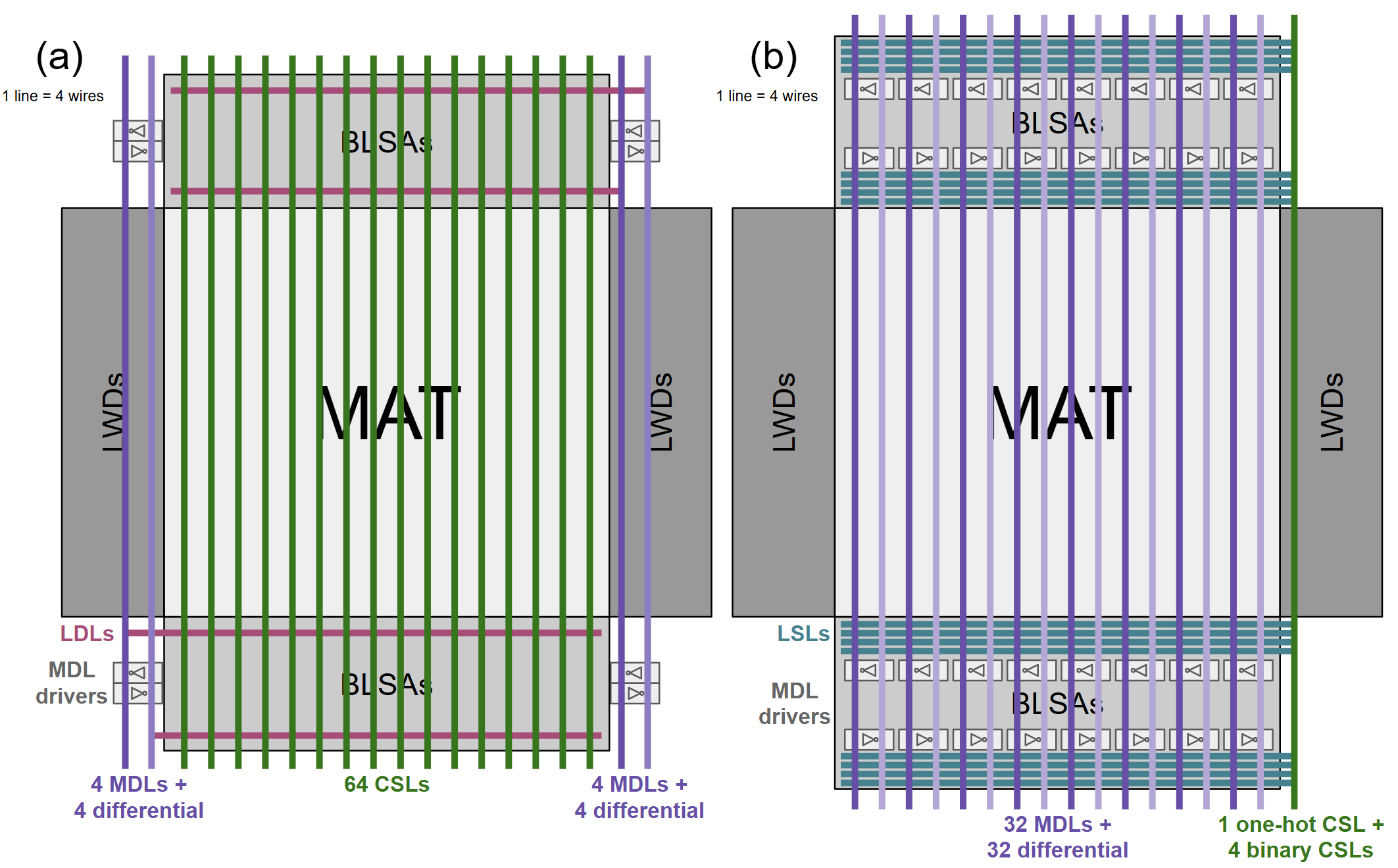}
    \caption{(a) A conventional MAT. (b) Proposed Dataline-over-MAT (DLOMAT). In both, 1 line represents 4 wires.}
    \label{fig:dlomat}
\end{figure}

\textbf{DreamRAM MAT Parameterization.}
DreamRAM adopts MAT-level metal layers and routing from \cite{chatterjee_architecting_2017}\cite{marazzi_hifi-dram_2024} as shown in Fig. \ref{fig:background} (d). At the MAT level, we parameterize the number of WLs and BLs per MAT, WL and BL isolation overheads, and MAT overhead for error correction (OD-ECC)\cite{ryu_16_2023}\cite{ha_understanding_2018}\cite{park_192-gb_2022}. For example, in the 512-BL MAT in Fig. \ref{fig:dlomat} (a), 1 of 64 CSLs selects 8 BLs to connect to 8 LDLs and 8 MDLs shared between MATs in the BL direction. One-hot CSLs occupy the wide cell array, while MDLs are restricted to the narrow LWDs. We introduce several parameters that help unlock more bandwidth per MAT. In conventional MATs (Fig. \ref{fig:dlomat} (a)), we parameterize the number of LDLs and MDLs per MAT, where more MDLs directly increase MAT bandwidth but introduce wiring area overhead over the BLSA and LWD. While DRAM density is important, DreamRAM opens this aspect of the design space and exposes the area vs. bandwidth tradeoff to the user. We allow designs with fewer MDLs per page (e.g., due to partial pages) to fire multiple CSLs back-to-back to obtain a full atom, as in \cite{chatterjee_architecting_2017}\cite{oconnor_fine-grained_2017}, which we term ``multi-pump.'' Pumps extend tCCDL, as explained in Section \ref{sec:corefreq}, but can be compensated for through the independence of the partial pages\cite{oconnor_fine-grained_2017}. Note that only dividing the MATs into smaller pages \cite{chatterjee_architecting_2017}\cite{ha_improving_2016} improves activation energy but does not increase MAT bandwidth. 

\textbf{Dataline-over-MAT (DLOMAT).} We propose a new MAT routing scheme, DLOMAT, depicted in Fig. \ref{fig:dlomat} (b). In DLOMAT, a parameterized number of MDLs are routed over the cell array where the CSLs were. For example, if we input 32 MDLs in a 512-BL MAT, we get 16 CSLs, which we route over the LWDs where the MDLs were. The former LDLs are now ``local select lines'' (LSLs) to route the CSLs to the BLSAs, swapping the connections of the BLSA column select transistors. The MDL drivers must move from beside the BLSA to be inside the BLSAs, though the height overhead is amortized across 8 BLSAs, while providing more routing space for LSLs. To minimize routing, we convert most of the CSLs from one-hot to binary for fewer CSLs per LWD. We keep a single one-hot CSL per pump for speed, where the rest form an extension of the column address and do not change during multi-pumping of multiple one-hot CSLs. We place repeaters and decode circuits where the MDL drivers were. We expect DLOMAT to raise per-MAT bandwidth, creating opportunities to trade off bandwidth with other design metrics.

\subsection{Timing Estimation}\label{sec:corefreq}
We take baseline timing parameters and scale them based on capacitance, resistance, pitch, and length. We assume wire delay scales linearly with length due to repeaters and flop circuits. We use capacitances from \cite{ha_understanding_2018}, which include drivers.

The \textbf{row miss latency} is defined as $\text{tRP}+\text{tRCD}+\text{tCL}$ (precharge$+$activate$+$read). We take baseline BLSA timing from \cite{kim_sensing_2019}.
We separate each of \textbf{tRCD} and \textbf{tRP} into signal propagation and bitline amplification portions: the signal portion scales with the bank width (farthest BLSA), while the bitline portion scales with $C_{cell}+n_{WL}C_{BL,\;per\;WL}+C_{BLSA}$. For OCSA's tRCD, the bitlines are separated from the sensing node during charge sense, during which we neglect $C_{BLSA}$ \cite{kim_sensing_2019}. For \textbf{tCL}, the worst-case read signal path runs from the base die edge to the TSVs, up to the top die, across to the DRAM die edge, and back; we scale tCL with $2n_{dies}R_{TSV}C_{TSV}+2l_{die,\;y}t_{die,\;per\;l}$. We extrapolate TSV parameters from \cite{weis_design_2011}.

The \textbf{bank cycle time} measures the maximum MDL rate, primarily limited by the RC constants of the CSL and MDL. It is scaled with $t_{CSL}+t_{LDL}+t_{MDL}+t_{MDL,PRE}+t_{DRV}$, where $t_{DRV}$ is a constant driver delay and the rest scale with their wires' capacitances. 
We calculate \textbf{tCCDL}, the delay between same-bank reads, as the product of bank cycle time and the number of pumps per atom. For ADL \cite{ryu_16_2023}, $\text{\textbf{tCCDS}}=\frac12 \text{tCCDL}$ is the delay between pseudo channel reads. We calculate bandwidth by tracing the databus widths and muxing from the MDLs out to DQs.

%% file: tab-validation.tex
\begin{table*}[!t]
\centering
\caption{Validation of DreamRAM Against Reported Measurements (Real / Model / Error)}
\begin{tabular}{l|
  r N r N r
  r N r N r
  r N r N r
  r N r N r
  r N r N r}
\toprule
\textbf{Validation Targets} 
& \multicolumn{5}{c}{\centering\textbf{Bandwidth (GB/s)}}
& \multicolumn{5}{c}{\centering\textbf{Capacity (GB)}}
& \multicolumn{5}{c}{\centering\textbf{Full- \& Closed-Row Energy (pJ/b)}}
& \multicolumn{5}{c}{\centering\textbf{Miss Latency (ns)}}
& \multicolumn{5}{c}{\centering\textbf{Die Area (mm$^2$)}} \\
\midrule
HBM3\cite{ryu_16_2023}
& \multicolumn{5}{r}{1024 / \text{1024} / 0.0\%}
& \multicolumn{5}{c}{16 / \text{16} / 0.0\%}
& \multicolumn{5}{c}{NA / \text{0.98--3.01} / NA\%}
& \multicolumn{5}{c}{NA / \text{64.2} / NA\%}
& \multicolumn{5}{c}{121 / \text{111.0} / -8.3\%} \\
HBM2E\cite{chun_16-gb_2021}
& \multicolumn{5}{r}{640 / \text{741} / 15.7\%}
& \multicolumn{5}{c}{16 / \text{16} / 0.0\%}
& \multicolumn{5}{c}{NA / \text{1.46--3.61} / NA\%}
& \multicolumn{5}{c}{NA / \text{61.1} / NA\%}
& \multicolumn{5}{c}{110 / \text{109.3} / -0.6\%} \\
\bottomrule
\end{tabular}
\label{tab:validation}
\end{table*}

%% file: tab-volume5d.tex
\begin{table*}[!t]
\centering
\caption{Summary of Design Space Volume (Relative to Full Design Space) and Ranges (Relative to Baseline)}
\label{tab:5d-volume-endash-centering}
\begin{tabular}{l
  l N r
  l N r
  l N r
  l N r
  l N r}
\toprule
Design Space
& \multicolumn{3}{c}{A}
& \multicolumn{3}{c}{B}
& \multicolumn{3}{c}{C}
& \multicolumn{3}{c}{D}
& \multicolumn{3}{c}{E} \\
\midrule
Number of Designs
& \multicolumn{3}{c}{205}
& \multicolumn{3}{c}{5766}
& \multicolumn{3}{c}{93285}
& \multicolumn{3}{c}{1409312}
& \multicolumn{3}{c}{2762776} \\
Convex Hull Volume (\%)
& \multicolumn{3}{c}{$0.0001\%$}
& \multicolumn{3}{c}{$0.006\%$}
& \multicolumn{3}{c}{$2.1\%$}
& \multicolumn{3}{c}{$75.3\%$}
& \multicolumn{3}{c}{$100\%$} \\
\midrule
Bandwidth (GB/s)
 & 0.50$\times$ & \sep{--} & \textbf{4.00$\times$} 
 & 0.18$\times$ & \sep{--} & \textbf{7.42$\times$} 
 & 0.05$\times$ & \sep{--} & \textbf{7.42$\times$} 
 & 0.04$\times$ & \sep{--} & \textbf{11.82$\times$} 
 & 0.04$\times$ & \sep{--} & \textbf{13.38$\times$} \\
Capacity (GB)
 & 0.13$\times$ & \sep{--} & \textbf{2.00$\times$} 
 & 0.03$\times$ & \sep{--} & \textbf{2.00$\times$} 
 & 0.02$\times$ & \sep{--} & \textbf{2.00$\times$} 
 & 0.01$\times$ & \sep{--} & \textbf{4.50$\times$} 
 & 0.01$\times$ & \sep{--} & \textbf{4.50$\times$} \\
Closed-Row Energy (pJ/b)
 & \textbf{0.76$\times$} & \sep{--} & 1.25$\times$ 
 & \textbf{0.40$\times$} & \sep{--} & 1.28$\times$ 
 & \textbf{0.16$\times$} & \sep{--} & 1.31$\times$ 
 & \textbf{0.11$\times$} & \sep{--} & 2.64$\times$ 
 & \textbf{0.11$\times$} & \sep{--} & 2.77$\times$ \\
Miss Latency (ns)
 & \textbf{0.59$\times$} & \sep{--} & 2.12$\times$
 & \textbf{0.55$\times$} & \sep{--} & 2.25$\times$
 & \textbf{0.51$\times$} & \sep{--} & 2.30$\times$
 & \textbf{0.39$\times$} & \sep{--} & 9.63$\times$
 & \textbf{0.38$\times$} & \sep{--} & 11.37$\times$ \\
Total Area (mm$^2$)
 & \textbf{0.16$\times$} & \sep{--} & 2.11$\times$
 & \textbf{0.07$\times$} & \sep{--} & 2.44$\times$ 
 & \textbf{0.04$\times$} & \sep{--} & 2.49$\times$
 & \textbf{0.03$\times$} & \sep{--} & 2.87$\times$ 
 & \textbf{0.03$\times$} & \sep{--} & 2.87$\times$ \\
\bottomrule
\end{tabular}
\end{table*}

%% file: 06results.tex
\section{Validation and Results}\label{sec:results}
We validate DreamRAM against measurements reported by industry HBM3 and HBM2E \cite{ryu_16_2023}\cite{chun_16-gb_2021} as shown in Table \ref{tab:validation}. We describe each design in DreamRAM parameters, and for HBM2E\cite{chun_16-gb_2021}, we set the scaled technology node to the reported 1ynm with the reported voltage domains, but do not modify node scaling confidence parameters. DreamRAM reports energy per bit as both (1) best-case ``full-row'' access, with whole pages accessed per ACT, and (2) worst-case ``closed-row'' policy, with only one atom accessed per ACT. 
While \cite{ryu_16_2023}\cite{chun_16-gb_2021} do not report energy or latency, the thorough parameterization of DreamRAM allows us to estimate these metrics across designs.

\subsection{Design Space Tiers, Sweep, and Visualization}
DreamRAM showcases a large range of design knobs with fine-grained parameters at all levels of the DRAM, from straightforward organizational knobs to granular routing reworks. Recognizing that not every parameter in DreamRAM is accessible to all designers, we divide DreamRAM's parameters into five tiers from most to least constrained (A--E), adding increasingly fine-grained parameters. Each tier contains the design points of the previous tiers. 

\setcounter{figure}{4}
\begin{figure*}[!t]
  \centering
  \begin{subfigure}{0.48\textwidth}
    \centering
    \includegraphics[width=\linewidth]{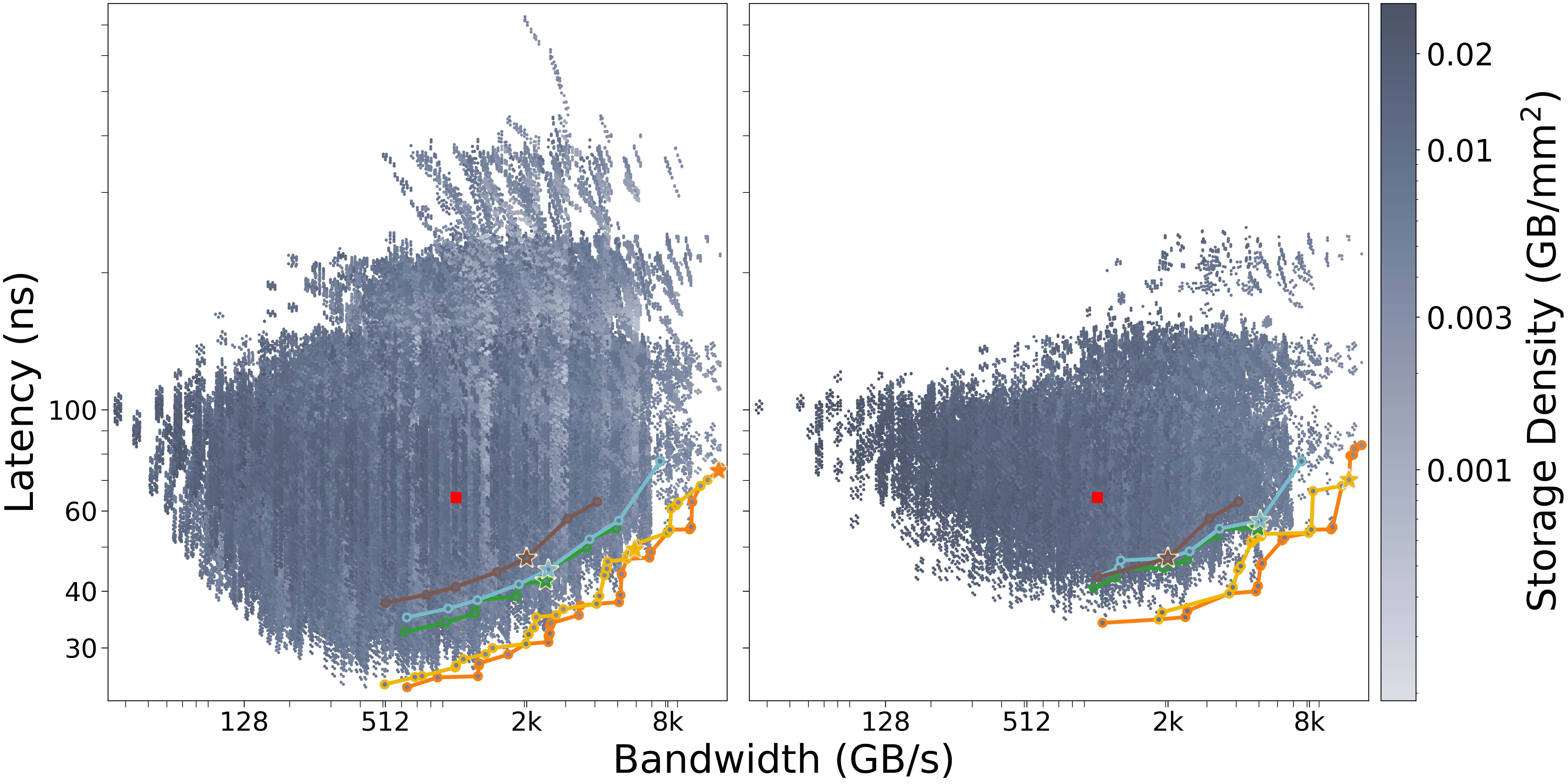}
    \subcaption{\textbf{Server CPU}.}
  \end{subfigure}
  \hspace{0.02\textwidth}
  \begin{subfigure}{0.48\textwidth}
    \centering
    \includegraphics[width=\linewidth]{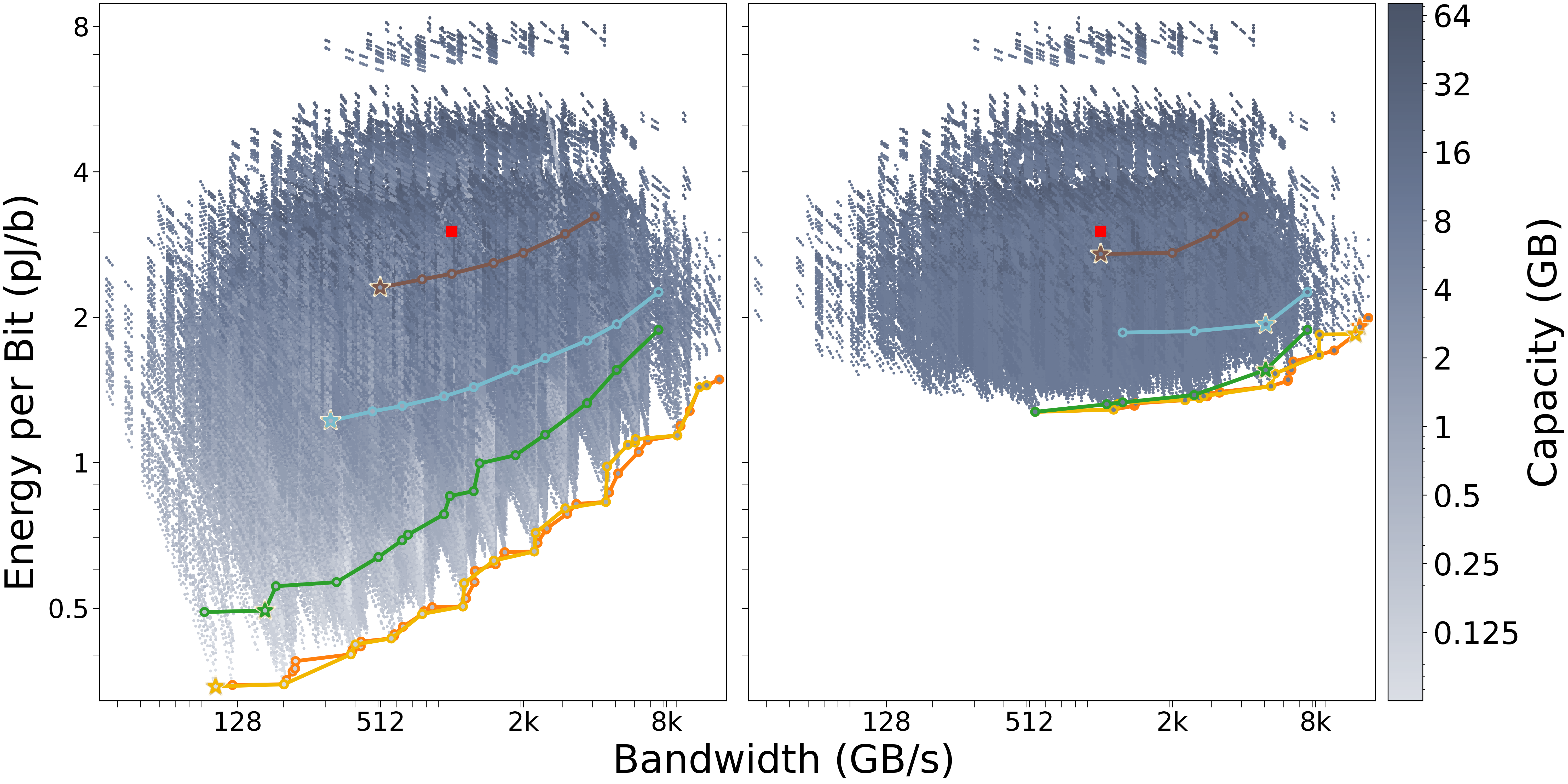}
    \subcaption{\textbf{Server GPU}}
  \end{subfigure}

  \vspace{0.01\textheight}

  \begin{subfigure}{0.48\textwidth}
    \centering
    \includegraphics[width=\linewidth]{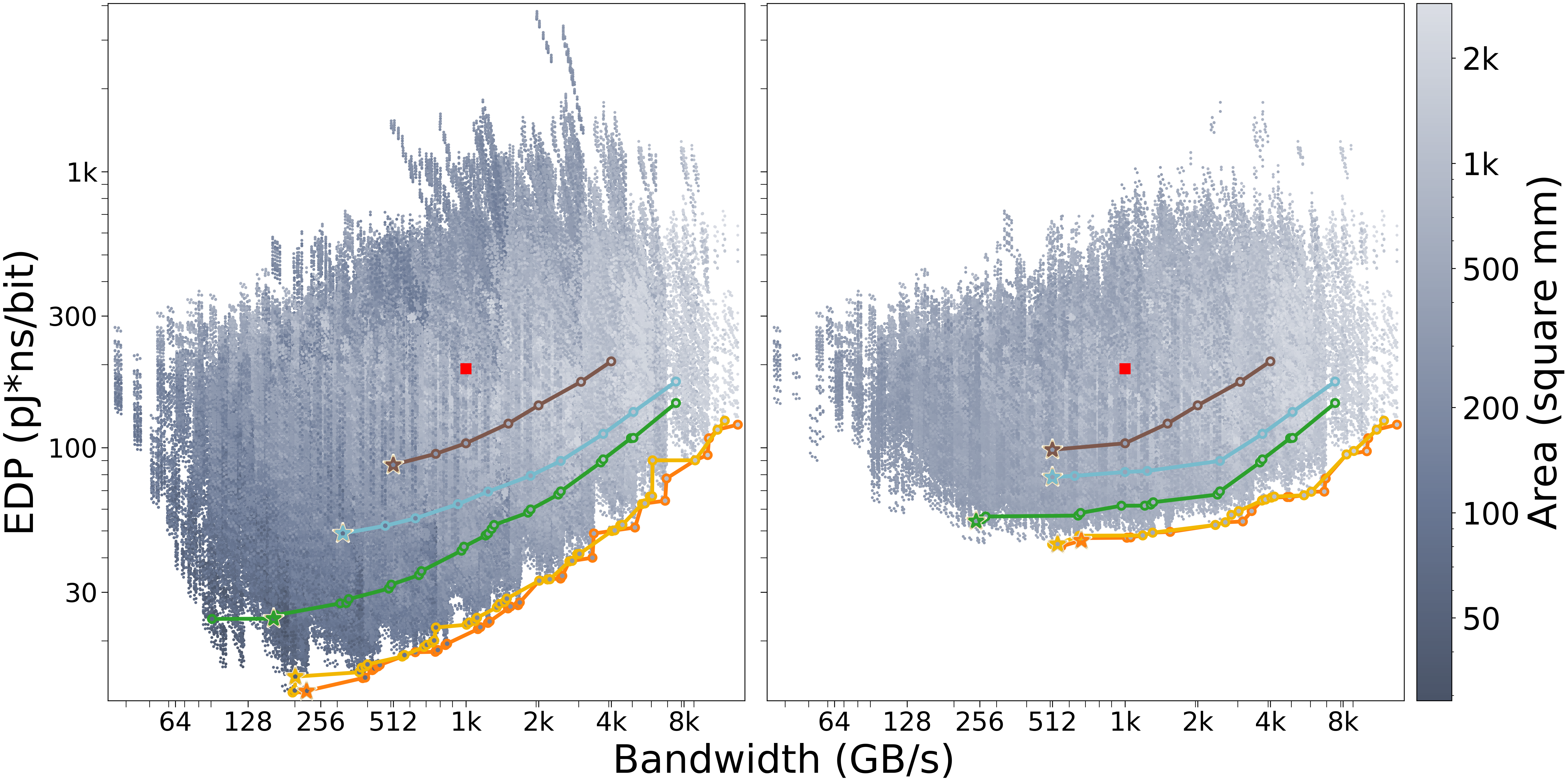}
    \subcaption{\textbf{High Performance Edge}}
  \end{subfigure}
  \hspace{0.02\textwidth}
  \begin{subfigure}{0.48\textwidth}
    \centering
    \includegraphics[width=\linewidth]{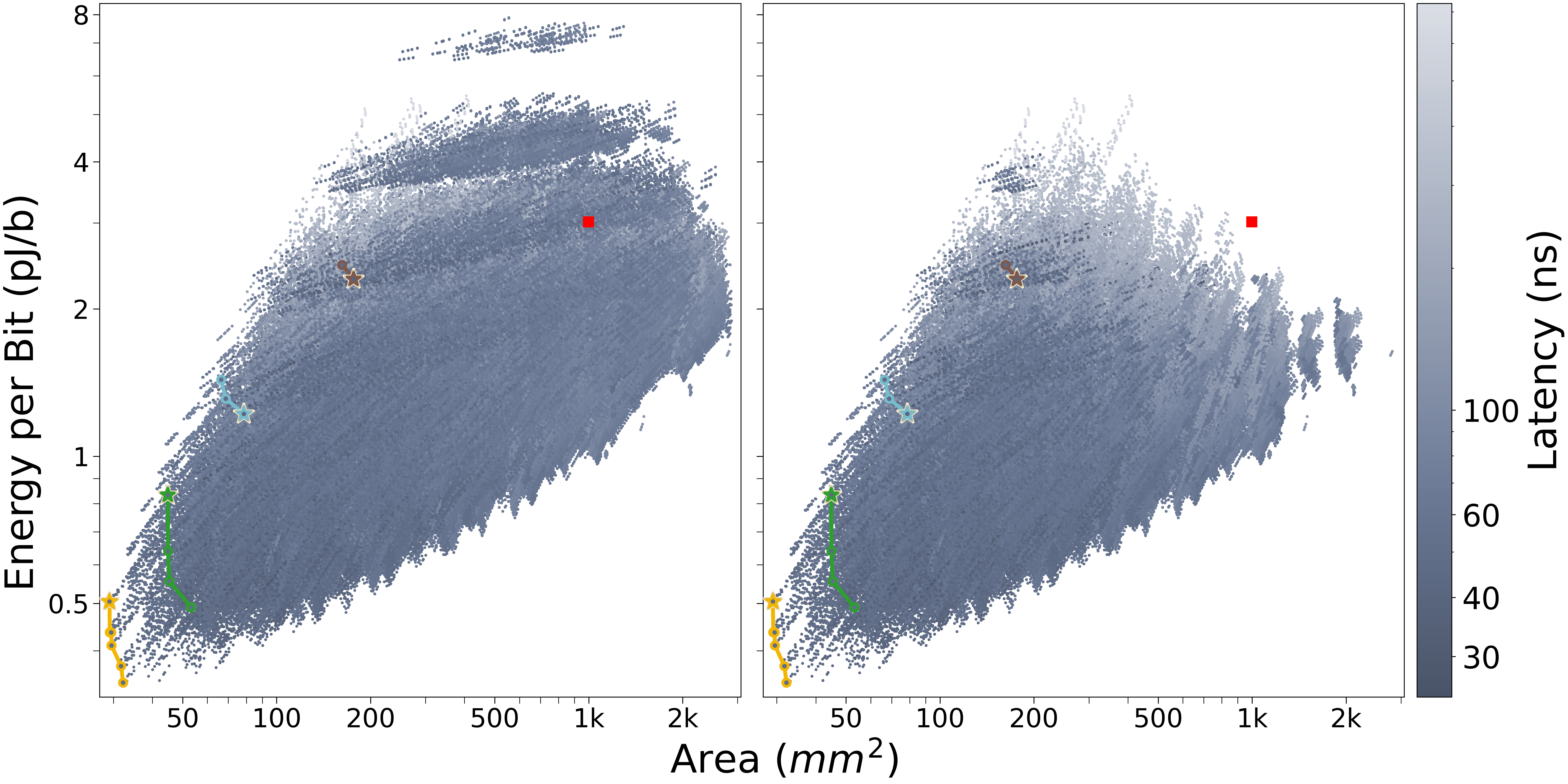}
    \subcaption{\textbf{Embedded IoT}}
  \end{subfigure}

  \caption{Paretos for each tier per application scenario. The left plot depicts the full design space, while the right plot is capacity-filtered per application. As additional parameters become available in each tier, the frontier shifts monotonically toward the ideal.}
  \label{fig:results}
\end{figure*}

\begin{itemize}
    \item \textcolor{Brown}{\textbf{Tier A: }}\textbf{Inter-bank level only}. Numbers of ranks, channels, channels per die, bank groups, banks, BGBUS mux
\item \textcolor{Cerulean}{\textbf{Tier B: }}\textbf{Bank level}. Tier A + number of subarrays per bank, number of MATs per subarray, SALP \cite{kim_case_2012}\footnote{SALP only modifies the bank row decoder, not subarray structures.}
\item \textcolor{Green}{\textbf{Tier C: }}\textbf{Subarray level}. Tier B + partial-page proposals\cite{chatterjee_architecting_2017}\cite{ha_improving_2016}, OCSA \cite{kim_sensing_2019}
\item \textcolor{Dandelion}{\textbf{Tier D: }}\textbf{MAT level}. Tier C + numbers of BLs, WLs, and MDLs per MAT
\item \textcolor{orange}{\textbf{Tier E: }}\textbf{Full Design Space}. Tier D + DLOMAT
\end{itemize}

We run a sweep of all these design parameters, with values above and below baseline where possible, and record each design's tier. We discard designs with stack height over 16 dies and die length or width over 13 mm, slightly larger than current HBMs \cite{ryu_16_2023}\cite{park_192-gb_2022}\cite{chun_16-gb_2021}. This yields a sweep of 2.8M design configurations used for all figures and results. 

Table \ref{tab:5d-volume-endash-centering} quantifies the reach of each tier, with ranges relative to baseline. Percent convex hull volume of a tier is the 5D (bandwidth, capacity, energy, latency, area) volume of the convex hull of the tier's points, normalized by that of the full design space (tier E). Each tier exposes a significantly larger range of designs that were inaccessible to the previous tier. Fig.~\ref{fig:design_space} showcases one projection of the 5D design space onto the 2D plane, with the tiers colored. 
\setcounter{figure}{3}
\begin{figure}[ht]
    \centering
    \includegraphics[width=0.9\linewidth]{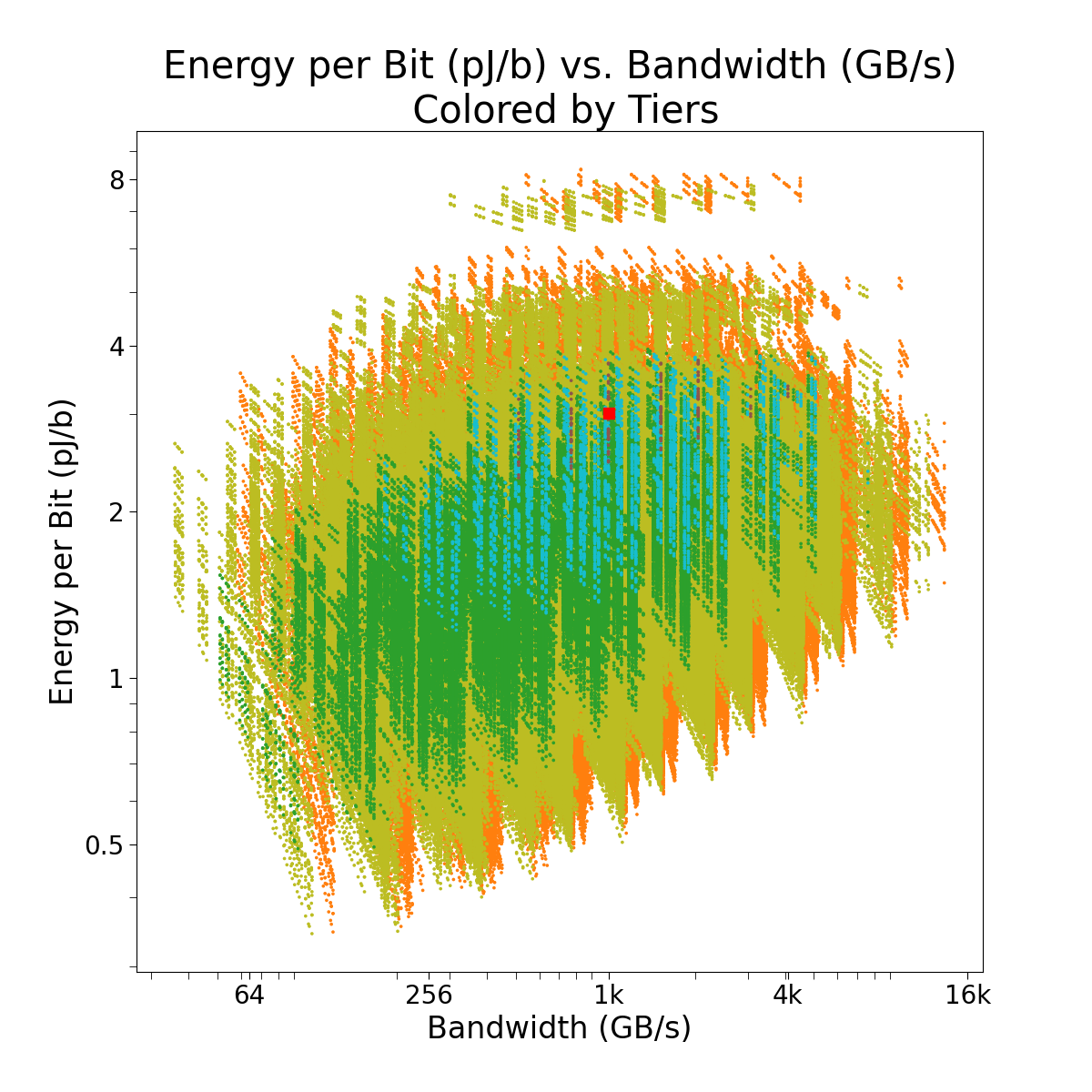}
    \caption{Projection of design space onto energy and bandwidth, colored by tiers. The square is the DreamRAM baseline HBM3.}
    \label{fig:design_space}
\end{figure}

We analyze the input parameters of the best-performing designs. Generally, designs with \textbf{high capacity} have more physical structures (dies, subarrays, MATs, etc.); designs with \textbf{low area} have fewer physical structures; designs with \textbf{high bandwidth} have more channels/DQs, higher core frequency (mostly by reducing bank height) and expose more datalines (e.g., DLOMAT); designs with \textbf{low latency} have short wires/TSVs; and designs with \textbf{low energy per bit} (closed-row) have short wires/TSVs and small pages (narrower banks or partial pages). With an understanding of each metric's preferred input parameters, we next analyze how to trade off between these metrics.

\input{tab-simulators-comparison}

\subsection{Analysis of Application Design Scenarios}\label{sec:scenarios}
We consider four application scenarios that impose differing demands on DRAM systems. For each scenario, we select two primary metrics and one secondary metric. In Fig. \ref{fig:results}, the two primary metrics define a Pareto frontier per tier, while the secondary metric is color-mapped and used to further differentiate 
among Pareto-equivalent points. The left plot presents the full design space, while the right plot applies application-based capacity filters. For clear visualization, we constrain our analysis to these 2D-plus-color projections of the 5D space; more complex objectives and tradeoffs with additional filters or metrics are possible for more comprehensive design studies.

\noindent\textbf{Server CPU} (Fig. \ref{fig:results} (a)). Traditional servers\cite{noauthor_leading_nodate}\cite{noauthor_intel_nodate}
house and run a large number of latency-sensitive workloads simultaneously. As such, performant systems must prioritize \textit{bandwidth and latency}. Rather than maximizing capacity or minimizing area in isolation, they target cost-aware storage density (capacity per unit area). We observe that Pareto designs tend towards shorter banks. While bandwidth prefers more channels/DQs, latency favors smaller dies and shorter stacks. 

\noindent\textbf{Server GPU} (Fig. \ref{fig:results} (b)). Server GPUs\cite{noauthor_nvidia_nodate}\cite{noauthor_nvidia_nodate-2} 
are generally throughput-driven machines that run applications with large data movement within fixed power envelopes. These systems primarily prioritize \textit{bandwidth and energy}. These applications require adequate capacity to operate. Pareto designs tend towards shorter banks, fewer bankgroups, and narrower MATs. While bandwidth prefers more channels and DQs, energy favors smaller pages, shorter banks, and shorter dies in the y direction (towards the TSV area). 

\noindent\textbf{High Performance Edge} (Fig. \ref{fig:results} (c)). High-performance edge systems \cite{noauthor_nvidia_nodate}\cite{noauthor_nvidia_nodate-1}\cite{noauthor_drive_nodate} for applications such as robotics, autonomous vehicles, and augmented reality operate on concurrent streams of data in real-time under tight power envelopes. Here, \textit{bandwidth, energy, and latency} are primary objectives; we capture energy and latency in the \textit{energy--delay product (EDP)}. These edge deployments are also subject to area constraints. Pareto designs tend towards shorter banks and narrower MATs. While bandwidth prefers more channels and DQs, EDP favors smaller pages, fewer dies, and shorter dies (y direction). 

\noindent\textbf{Embedded IoT} (Fig. \ref{fig:results} (d)). Embedded IoT systems\cite{ltd_buy_nodate}\cite{noauthor_imx_nodate}
perform periodic sensing, lightweight local processing, etc., under tight energy and area footprints. Within these physical constraints, these devices must respond to real-time events. So, these systems prioritize \textit{energy and area}, followed by latency. Pareto designs tend towards fewer physical structures including DQs, channels, banks, and MATs. While area favors larger unbroken pages, energy tends towards smaller pages and focuses more on reducing die size in the y direction. Although 3D die-stacking does not currently cater towards embedded edge devices, we envision a future of 3D integration where edge device logic can be added in or below the base die for a compact, well-performing solution. 

\subsection{Case Study: Optimizing HBM for Server GPUs}
\setcounter{figure}{5}
\begin{figure}
    \centering
    \includegraphics[width=1\linewidth]{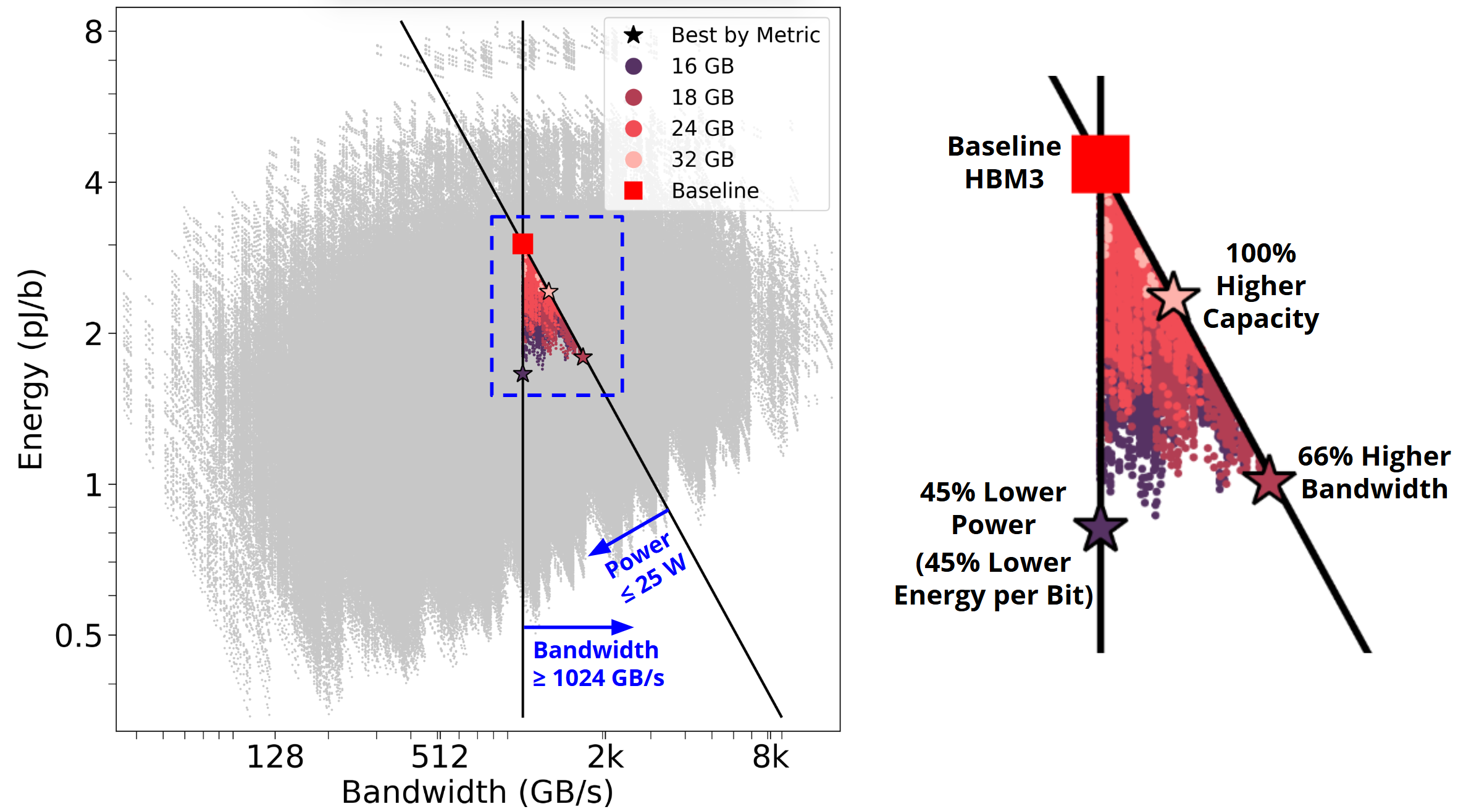}
    \caption{(a) Server GPU case study, requiring iso-capacity (color), iso-bandwidth (vertical line), and iso-power (diagonal line) compared to the baseline design. (b) Zoom-in showing the best designs for each metric.}
    \label{fig:abstract_results}
\end{figure}
We further demonstrate navigation of multi-dimensional constraints on DreamRAM's vast design space by optimizing server GPU memory metrics. Fig. \ref{fig:abstract_results} enforces additional constraints where designs must achieve bandwidth, capacity, and power no worse than the baseline (tiers are ignored). Under these constraints, we identify designs with each of 66\% higher bandwidth, 100\% higher capacity, and 45\% lower power and energy per bit. These best designs are consistent with the trends outlined in Section \ref{sec:scenarios}. Power optimization is similar to energy, but may sometimes slightly compromise energy per bit for a slower core frequency.

%% file: tab-simulators-comparison.tex
\begin{table*}[!ht]
\scriptsize
\centering
\caption{Comparison of DRAM Simulator Models and Output Capabilities}
\label{tab:dram-sim-comparison}
\begin{tabular}{l l c c c c c c}
\toprule
& 
& Controller models 
\cite{luo_ramulator_2023}\cite{noauthor_dramsys40_nodate}\cite{li_dramsim3_2020}
& CACTI-3DD\cite{ke_chen_cacti-3dd_2012}
& 3D-DATE\cite{park_3-d-date_2019}
& DRAMSpec\cite{weis_dramspec_2017}
& DRAMDSE\cite{ha_understanding_2018}
& \textbf{DreamRAM} \\ 
\midrule
\multirow{2}{*}{Classification} 
  & Simulator Type & Trace-Driven & Analytical & Analytical & Analytical & Analytical & Analytical\\
  & Open-Source     & \yes & \yes & \no & \yes & \no &  \yes\\
\midrule
\multirow{2}{*}{Parallelism} 
  & Pseudo Channels & \yes  & \no  & \no  & \no  & \no & \yes  \\
  & Bank Groups     &\yes & \no  & \no  & \no  & \yes & \yes  \\
\midrule
\multirow{4}{*}{Customization}
  & MAT Level  & \no & \no          & \no  & \yes & \yes & \yes   \\
  & Subarray Level & \no & \no      & \yes & \yes & \yes & \yes   \\
  & Bank Level & \yes & \no         & \yes & \yes & \yes & \yes   \\
  & Inter-Bank Level  & \yes & \yes & \no  & \no  & \yes & \yes   \\
\midrule
\multirow{4}{*}{Output Metrics} 
  & Bandwidth & \yes & \no & \no & \no & \no & \yes   \\
  & Capacity  & \no  & \no & \no & \no &\no & \yes \\
  & Energy/Power    & \yes & \yes & \yes & \no & \yes & \yes   \\
  & Latency   & \no & \no & \no & \yes & \no & \yes   \\
  & Area      & \no & \yes &\yes & \yes & \yes & \yes   \\
\bottomrule
\end{tabular}
\end{table*}

%% file: 07relatedworks.tex
\section{Related Works}\label{sec:relatedworks}

DreamRAM is compared to related models in Table \ref{tab:dram-sim-comparison}. Many existing DRAM simulators primarily model already-manufactured/profiled designs, offering limited ability to explore the design space of future memories. On the other hand, academic proposals often focus narrowly on specific techniques and design a single DRAM variant for that concept. We believe the performance of future computing systems will rely increasingly on their memory configuration, and we have built DreamRAM to enable us to open up and dissect the vast unrealized 3D die-stacked DRAM design space.

%% file: 08conclusion.tex
\section{Conclusion}

DreamRAM combines fine-grained parameterization with analytical modeling to expose and highlight the large custom DRAM design space across bandwidth, capacity, energy, latency, and area. 
The DreamRAM framework allows designers to evaluate tradeoffs, uncover new design opportunities, and integrate the development of next-generation systems with workload-tailored memories. 
We envision DreamRAM will enable applications to match with the memory system of their dreams.

\section*{Acknowledgements}
We gratefully acknowledge partial funding from NSF grant CCRI-2346435 and the Harvard College Research Program.

%% file: 88appendix.tex
\section*{Appendix}
\begin{figure*}[!htb]
    \centering
    \includegraphics[width=1\linewidth]{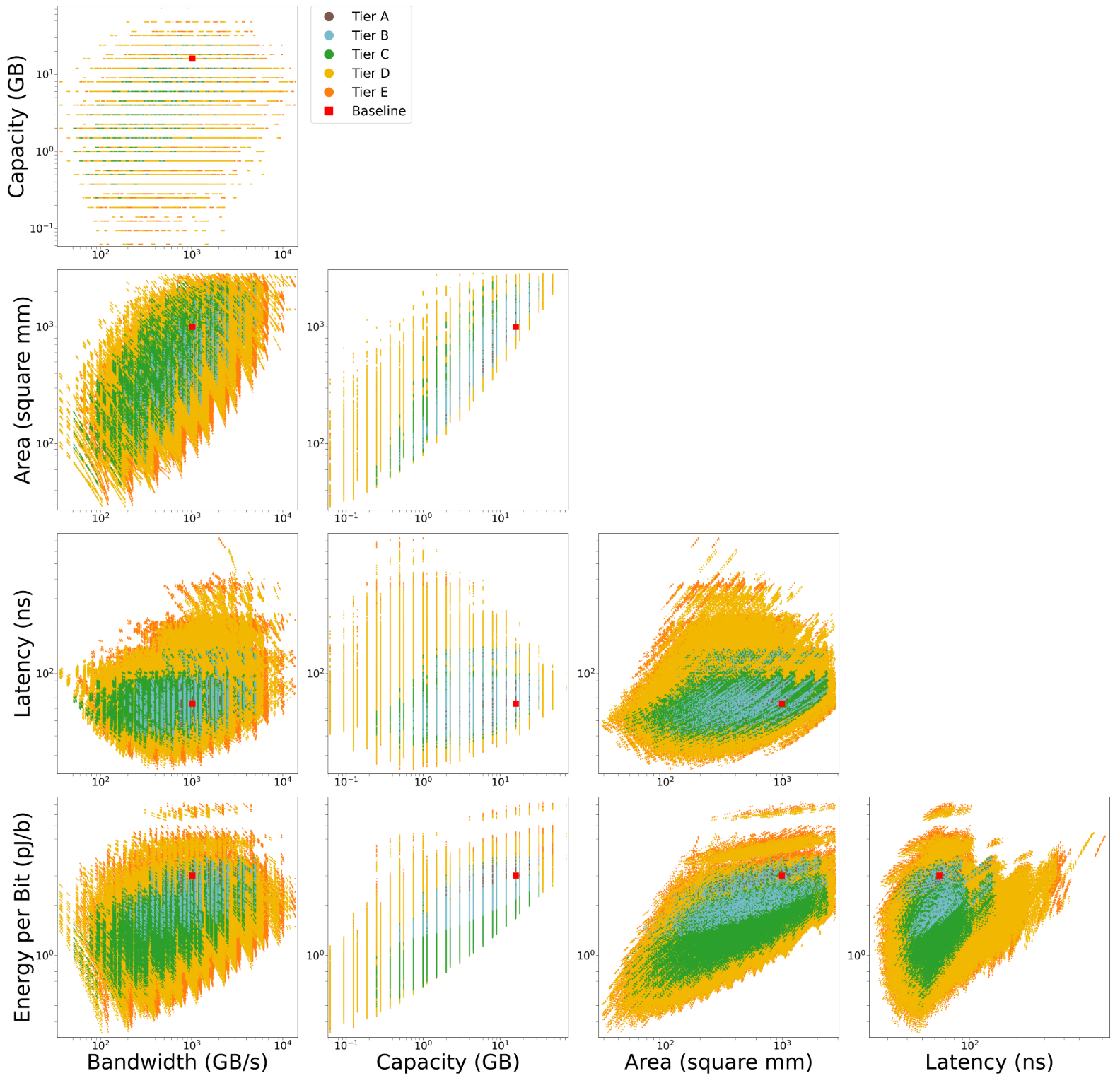}
    \caption{All $\genfrac{(}{)}{0pt}{}{5}{2}=10$ projections of the 5D design space onto two metrics at a time. Plots in any row or column share an axis; the axis labels for each row or column are along the left and bottom edges. The points are colored by tier, and the red square is the DreamRAM baseline HBM3.}
    \label{fig:5c2}
\end{figure*}

Fig. \ref{fig:5c2} shows all $\genfrac{(}{)}{0pt}{}{5}{2}=10$ 2D-projects of the 5D design space. Fig. \ref{fig:design_space} is one such projection, and is the energy/bit vs. bandwidth plot in the bottom left. The capacities are discretized by powers of 2 multiplied by 1, 3, or 9, which manifest as stripes whenever capacity is an axis. This is due to the sweep setup where the number of subarrays and the number of channels (and the number of channels per die) are each allowed to have a factor of 3. Having such a factor of 3 in the subarrays is already a common practice in commodity DRAM such as LPDDR4, and helps to fill in the gaps between discrete powers of 2. Of the metrics, area and capacity appear the most correlated, followed by energy/bit and capacity, followed by energy/bit and area. The main benefit of DLOMAT is its bandwidth boost of about 13\% for the highest bandwidth configurations vs. non-DLOMAT designs. 

As of publication, DreamRAM's focus was on modeling 3D die-stacked DRAMs through fine-grained parameterization at the inter-bank, bank, subarray, and MAT levels, unlocking a vast design space in DRAM bandwidth, capacity, energy, latency, and area. Parameterization at the smallest levels of the DRAM hierarchy is the most difficult, requiring finer circuit and device models. Cells are currently abstracted to their area and bitline capacitance. Cell- and BLSA-level characteristics, including retention time, RowHammer effects, read-to-precharge/row active time, warrant further modeling. 

Regarding RowHammer, in both DLOMAT and non-DLOMAT designs, the WL and BL structures under the cells are not modified. Since RowHammer is a row activation phenomenon, we posit that DreamRAM's modifications to the CSLs and MDLs for column accesses should not affect RowHammer susceptibility. DLOMAT's MDLs routed over the MAT are differential, which may cause less bounce than routing conventional single-ended CSLs over the MAT at the same frequency, though we make no claims about DLOMAT's data retention. ColumnDisturb is also a row activation phenomenon, even if the victim cells are along the columns.

DreamRAM designs vary significantly in their page size and numbers of rows, banks, and bank groups, and these are often the biggest drivers of the performance metrics. Still, precise bandwidth, latency, and energy will depend on how systems choose to take advantage of these DRAM design variations throughout DreamRAM's design space. 